\documentclass[traditabstract,rnote]{aa} 
\usepackage{txfonts}
\usepackage{times}
\usepackage{graphicx}
\usepackage{natbib}
\usepackage{mathptm}

\begin{document}

\title{Calibration update of the COMBO-17 CDFS catalogue}

\author{C. Wolf\inst{1} \and H. Hildebrandt\inst{2,3} E. N. Taylor\inst{3} \and K. Meisenheimer\inst{4}}
\institute{Department of Physics, University of Oxford, DWB, Keble Road, Oxford, OX1 3RH, U.K.; 
  \email{cwolf@astro.ox.ac.uk}
\and Argelander-Institut f\"ur Astronomie, Auf dem H\"ugel 71, D-53121 Bonn, Germany.  
\and Leiden Observatory, Leiden University, Niels Bohrweg 2, 2333CA Leiden, The Netherlands
\and Max-Planck-Institut f\"ur Astronomie, K\"onigstuhl 17, D-69117 Heidelberg, Germany.}

\date{Received;  accepted}

\abstract{
We present an update to the photometric calibration of the COMBO-17 catalogue on the Extended Chandra Deep Field South, which is now consistent with the GaBoDS and MUSYC catalogues. As a result, photometric redshifts become slightly more accurate, with $<0.01$ rms and little bias in the $\delta z/(1+z)$ of galaxies with $R<21$ and of QSOs with $R<24$. With increasing photon noise the rms of galaxies reaches 0.02 for $R<23$ and 0.035 at $R\approx 23.5$. Consequences for the rest-frame colours of galaxies at $z<1$ are discussed.
\keywords{Catalogs - Surveys - Techniques: photometric}}
\maketitle
\titlerunning{Calibration update of COMBO-17 CDFS }
\authorrunning{Wolf et al. }

\section{Introduction}

Almost five years ago, a catalogue of the Extended Chandra Deep Field South (ECDFS) was published by the COMBO-17 survey project \citep{Wolf04}. It comprised photometry in 17 bands, viz. the broad-band filters UBVRI and 12 medium-band filters covering the wavelength range from 400 to 930~nm. On the basis of the SEDs they derived and published photometric classifications into star, galaxy, QSO or white dwarf, as well as photometric redshifts for galaxies and QSOs. While the A901 and S11 fields of the COMBO-17 survey had straightforward calibrations, it was known that the ECDFS calibration was ambiguous. In spite of this, photometric redshifts in the ECDFS were of very good quality with $\sigma_z/(1+z)\approx 0.02$ for galaxies with $R<23$ and for QSOs (=type-1 AGN) brighter than $M_B<-22$. 

Recently, it became clear that the photometric calibration was indeed in error with an almost monotonic drift across wavelength. Since the publishing of the original COMBO-17 ECDFS catalogue, two separate projects -- MUSYC  \citep{MUS} and GaBoDS \citep{Gab} -- have constructed broadband photometric catalogues of the ECDFS. The GaBoDS consortium retrieved and combined all existing WFI imaging (up to Dec 2005), including raw COMBO-17 data, and calibrated them from nightly standard star imaging. Both the GaBoDS and MUSYC projects use the same WFI data; the MUSYC catalogue adds NIR data to the GaBoDS--reduced optical data. It was through comparison of the catalogue photometry between them and COMBO-17 that the calibration issue was first noticed.

Broad-band photo-z's are susceptible to calibration offsets, but photo-z's from medium-band surveys are more robust against calibration errors. This applies to both random calibration offsets, whose impact is diminished by a large number of bands, and to systematic calibration drifts, since spectral features are reliably located by medium-band filters even when the SED is globally wrong. Hence, the presence of the calibration drift could not be deduced from the fully satisfying photo-z performance.

In this paper, we publish a refined calibration alongside an updated catalogue and explain the origin of the calibration error (see Sect.~2), the details of which might be useful for future multi-band surveys. We then compare the new calibration to the two other ground-based surveys that quantified the calibration difference in the broad-bands originally (see Sect.~3). In Sect.~4 we discuss subtle consequences for the photo-z accuracy and the more important update of the rest-frame colours.

\section{The CDFS calibration issue and update}

The COMBO-17 survey has been calibrated with the same technique as the CADIS survey \citep{CAD}: Both surveys have a large number of filters, collect photons irrespective of sky transparency and react to changing seeing conditions by scheduling the observations in different filters very flexibly. It was thus desired to conduct the survey imaging free from the need to obtain regular standard star calibration imaging, which would produce very large overheads with a large number of filters. This approach is possible when spectrophotometric standard stars are available inside the survey imaging fields, since then every survey exposure contains a standard star. Both surveys established two spectrophotometric standards in each of their fields with a separate spectroscopy programme in photometric nights.

This approach worked very satisfyingly for the seven CADIS fields and for all COMBO-17 fields except for the CDFS: here, the two standard stars displayed a calibration difference drifting monotonically with wavelength over the entire range from U-band to I-band, a phenomenon that can not be caused e.g. by missing order separation filters. Clearly, one of the two standards in the CDFS had a true spectrum that differed from the one obtained in the calibration spectroscopy. The reason for this is still unknown: The calibration spectroscopy used a $5\arcsec$ wide slit in a seeing of $1\arcsec$ so that any non-optimal centring of the star can not produce the observed calibration drifts. 

We have considered two hypotheses: (i) the wrong star was observed by accident, however, a review of the observing data seems to rule this out; (ii) the star with the wrong calibration is an eclipsing binary undergoing an eclipse at the time of spectroscopy; the standard stars show no sign of variability during {\it all} of the survey imaging, so this explanation may sound unlikely, but we can not rule it out.

\begin{table}
\caption{Calibration offsets resulting from changing the standard star: to be added to the CDFS photometry in Wolf et al. (2004). \label{recal}}
\begin{tabular}{lc|lc|lc}
\hline \noalign{\smallskip} \hline \noalign{\smallskip} 
filter	&  $\Delta m$ & filter	&  $\Delta m$  & filter	&  $\Delta m$  \\
\noalign{\smallskip} \hline
U	& $-0.143$	& 420	& $+0.038$	& 646	& $-0.022$\\
B	& $+0.040$	& 464	& $-0.027$	& 696	& $-0.081$\\
V	& $+0.003$	& 485	& $+0.037$	& 753	& $-0.092$\\
R	& $-0.054$	& 518	& $+0.039$	& 815	& $-0.103$\\
I	& $-0.123$	& 571	& $-0.033$	& 855	& $-0.146$\\
 	&  			& 604	& $-0.072$	& 915	& $-0.144$\\
\noalign{\smallskip} \hline
\end{tabular}
\end{table}

The ideal path to take in such circumstances is repeating the calibration and accepting ensuing delays. The path that was taken in practice was trying to identify which star was faulty and proceed with only one calibration standard. The COMBO-17 procedure for reviewing the calibration included comparing the colours of bright ($R<20$) point sources with those of a library of stellar spectra \citep{P98}; usually, differences could be traced back to problems in the raw image reduction of an affected filter. In the CDFS data no particular filter had processing faults, but the two standard stars still disagreed with the overall Pickles library to a small and similar extent in opposite directions; a decision in favour of one star was taken.

The calibration update presented here is entirely a result of changing from one standard star to the other, and the new choice has been consistently supported by three pieces of evidence: 

\begin{itemize}
\item The colour differences between the initial ECDFS calibration of COMBO-17 and the independently calibrated GaBoDS and MUSYC catalogues vanished into insignificance after choosing the alternative standard star.
\item During the survey calibration for HIROCS (Heidelberg IR-Optical Cluster Survey) it was realised that the stars found in typical extragalactic fields are predominantly main-sequence stars, while the Pickles atlas also contains many subgiants and giants whose colours differ from those of main-sequence stars; when the colours of the COMBO-17 point sources were then compared to only main-sequence Pickles stars, the stellar locus narrowed and it became immediately clear which of the two COMBO-17 standards was faulty.
\item Photometric redshifts obtained from broad-band filters alone showed biases in the original photometry and template repair methods suggested offsets on the order of those applied \cite{HWB08}.
\end{itemize}

\begin{figure}
\centering
\includegraphics[clip,width=0.87\hsize]{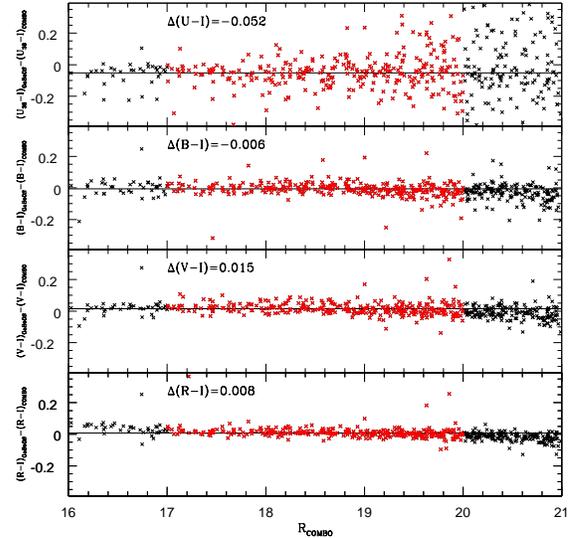}
\caption{Comparison of new COMBO-17 and GaBoDS photometry. Now, the SED shapes of COMBO-17 and GaBoDS agree. An offset of $I_{\rm C17}-I_{\rm Gab}=0\fm077$ or $I_{\rm C17}-I_{\rm Mus}\approx 0\fm10$ remains. \label{calibcomp}}
\end{figure}

In Table~\ref{recal} we list magnitude values to be added to the initial calibration in order to obtain the refined version. Negative numbers mean that the new calibration renders objects brighter. The overall trend is to make the SEDs redder than before. The updated catalogue can also be obtained in FITS or ASCII format from the COMBO-17 website\footnote{http://www.mpia.de/COMBO/combo\_CDFSpublic.html} and in ASCII only from CDS\footnote{http://cdsweb.u-strasbg.fr/CDS.html}. We also flagged a number of additional objects, whose total photometry is affected by close neighbours and remind the reader only to rely on the photometry with objects having $flag\_all<8$.

\begin{figure*}
\centering
\includegraphics[clip,angle=270,width=0.8\hsize]{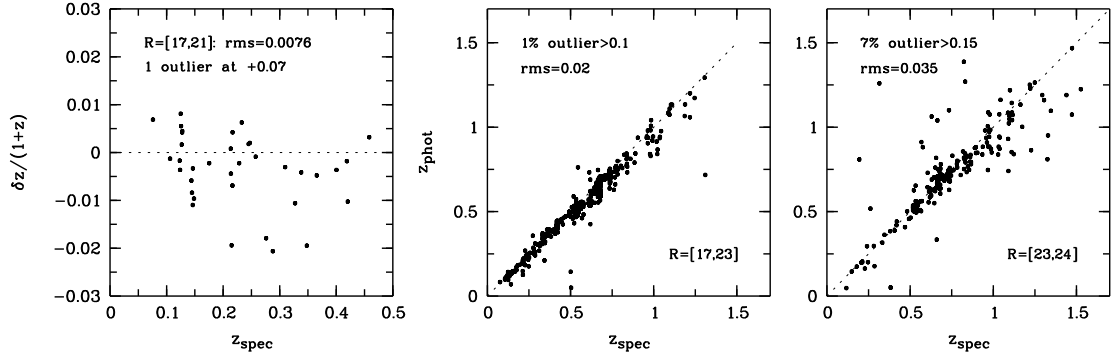}
\caption{Comparison of photometric and spectroscopic galaxy redshifts: precision is $<0.01$ rms at $R<21$ but changes with brightness. \label{speccomp}}
\end{figure*}

\section{Calibration comparisons; aperture photometry}

We compare the new COMBO-17 calibration with the broad-band photometry of GaBoDS in Fig.~1, demonstrating good agreement in terms of the colours. Except for the notoriously difficult U-band, colour indices are within $\sim 0\fm01$ of each other. The overall calibration of the BVRI filters shows a difference of $\la 0\fm1$, such that objects in COMBO-17 appear to be fainter. The calibration of MUSYC and GaBoDS agrees very well with that of the GOODS data set \citep{MUS}.

This comparison covers only point sources, because multi-colour data are provided in the COMBO-17 catalogue only for aperture photometry but not for total photometry. In the presence of colour gradients, which are common in galaxies, total colours are not expected to be identical to aperture colours. The only total photometry in COMBO-17 is in the R-band, which is used for the normalisation of the SED to obtain luminosities and masses. Here, we briefly repeat the unchanged principles behind the COMBO-17 aperture SEDs:

We obtained spectral energy distributions of all objects from photometry in all 17 passbands by projecting the known object coordinates into the frames of reference of each single exposure and measuring the object fluxes at the given locations. In order to optimize the signal-to-noise ratio, we measure the spectral shape in the high surface brightness regions of the objects and ignore potential low surface brightness features at large distance from the center. However, this implies that for large galaxies at low redshifts $z<0.2$ we measure the SED of the central region and ignore colour gradients.

Also, we suppressed the propagation of seeing variations into the photometry by making sure that we always probe the same physical footprint outside the atmosphere $f(x,y)$ of any object in all bands irrespective of the PSF $p(x,y)$. Here, the footprint $f(x,y)$ is the convolution of the PSF $p(x,y)$ with the aperture weighting function $a(x,y)$. If all three are Gaussians, an identical physical footprint can be probed even when the PSF changes, simply by adjusting the weighting function $a(x,y)$. We chose 
to measure fluxes on a footprint of $1\farcs5$ FWHM outside the atmosphere. In practise, we use the package MPIAPHOT to measure the PSF on individual frames, choose the weighting function needed to conserve the footprint and obtain the flux on the footprint \citep{MR93}. Fluxes of individual frames are averaged and the flux error is derived from the scatter. Thus, it takes not only photon noise into account, but also suboptimal flatfielding and uncorrected CCD artifacts.

\section{Consequences for photometric classifications and redshifts}

A change in SED leads to changes in the most probable object class, estimated photometric redshift and restframe properties, at least formally. However, the new redshifts are very similar to the previous ones, and a comparison with spectroscopic samples shows that the new ones are moderately more accurate, mostly in the quality saturation regime at $R<21$. Medium-band photo-z's rely on locating spectral features between neighbouring medium-bands and are thus more robust against calibration offsets. Hence, the original CDFS photo-z's in COMBO-17 were already almost as precise as they could be. 

In contrast, broad-band photo-z's rely more on overall colours and calibration offsets translate directly into strong photo-z biases. Such biases were indeed found in a broad-band photo-z study \cite{HWB08} and were one of the arguments suggesting the recalibration of COMBO-17. In Hildebrandt et al. (2008) the published values of photo-z quality are already based on the corrected COMBO-17 photometry.

\begin{figure*}
\centering
\hbox{
\includegraphics[clip,angle=270,width=0.625\hsize]{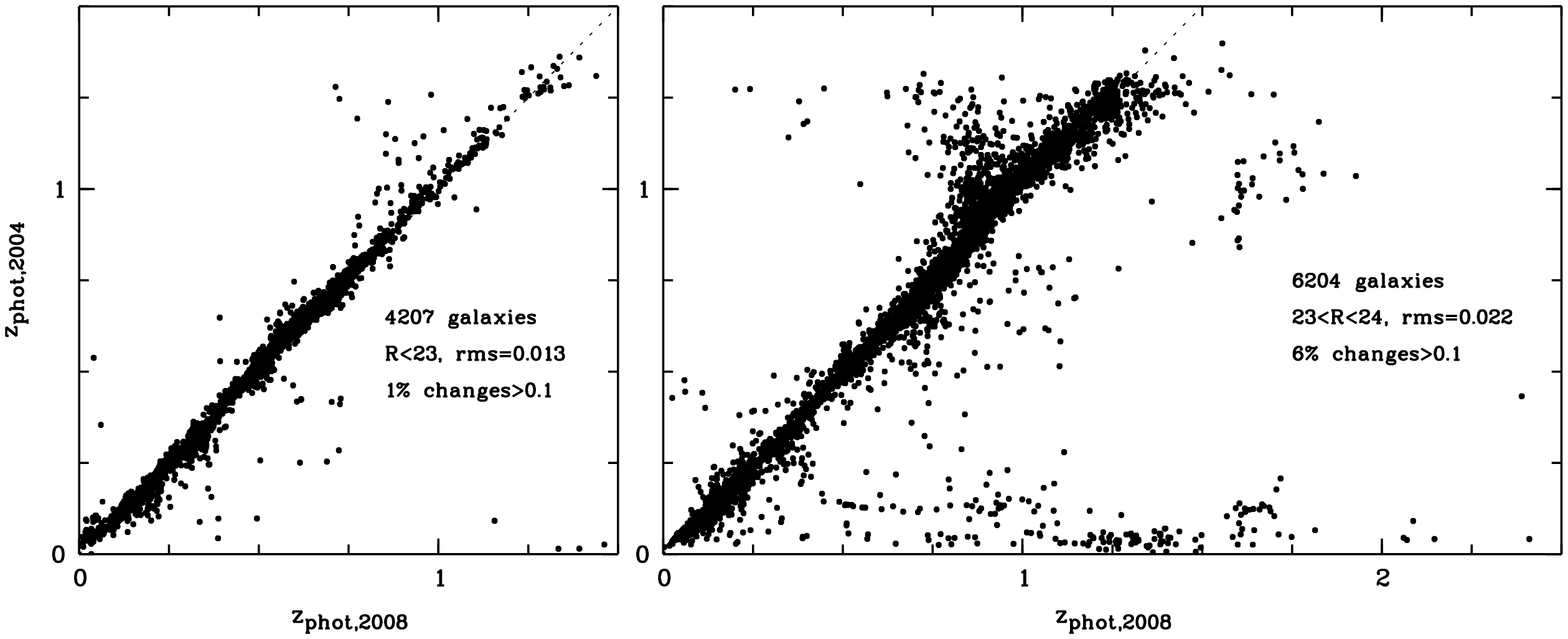}
\includegraphics[clip,angle=270,width=0.273\hsize]{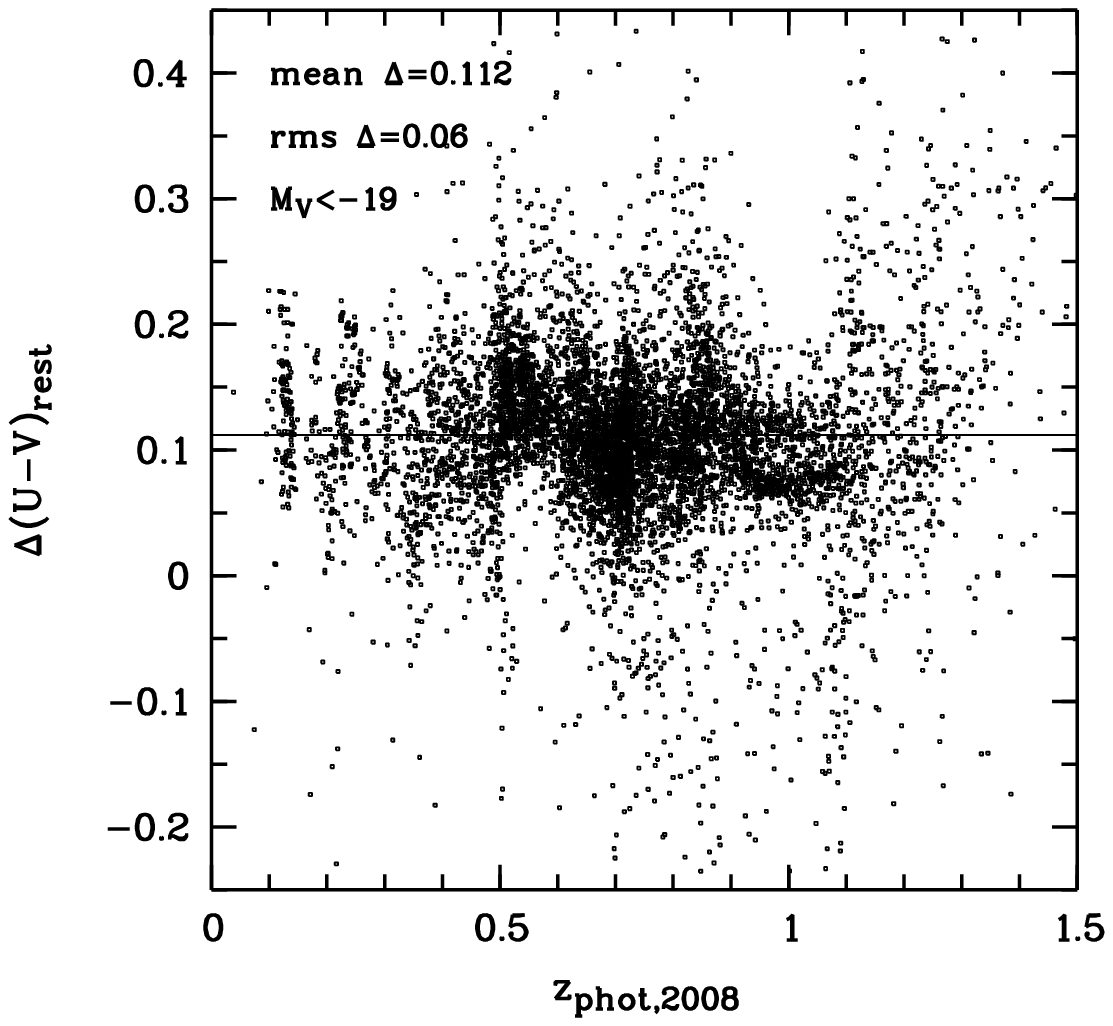}}
\caption{{\it Left and centre:} Comparison of old and new photometric redshifts. {\it Right:} Change in $U-V$ resulting from the new calibration. \label{zzcomp}}
\end{figure*}

\subsection{Galaxies}

On this occasion, we update the COMBO-17 photo-z procedure with two more changes: (i) we remove a bug in the treatment of bimodal redshift probability distributions, and (ii) we extend the redshift range under consideration for galaxies to $z\sim 7$ by default, although no galaxy brighter than $R=24$ ends up with a redshift estimate above $z=2.5$. In the following, we briefly discuss the resulting updated redshifts and classifications and restrict ourselves entirely to the magnitude range of $R<24$.

Fig.~\ref{speccomp} compares the updated galaxy photo-z's with the high-quality subset of the VVDS sample \citep{LeF04}. As with the initial COMBO-17 photo-z's on the CDFS the agreement is very good and the $\delta z/1(+z)$ deviations have an rms of $\sim 0.008$ at $R<21$, increasing to 0.02 at $R<23$ and $\sim 0.035$ at $R=[23,24]$ as photon noise widens the redshift probability distributions. At $R>24$ the signal in the medium-band filters is insufficient to provide an advantage over broad-band SEDs. These comparisons have effectively tested only the redshift range $z=[0,1.2]$, where an optical-only filter set is expected to perform well. We know little about the COMBO-17 photo-z accuracy of normal galaxies at $z>1.2$, but the reasoning behind not limiting the photo-z code to low redshifts any more is to avoid forcing a low-redshift solution onto a true high-redshift galaxy that would contaminate faint low-redshift samples.

Fig.~\ref{zzcomp} compares the galaxy redshifts and restframe colours of the initial data release from 2004 with this update. Redshifts have changed by $>0.1$ for about 1\% of the galaxies at $R<23$, but for more of the fainter galaxies, where the medium-band filters become less helpful in constraining spectral features, and the change of broad-band colours makes a larger difference. Most conspicuous is a horizontal feature of galaxies previously estimated to be at $z<0.2$, but now estimated to stretch from $\sim 0.5$ to beyond $z=1$. The new estimates for these objects are likely to be more correct than the previous low-z values, even though we do not know how accurate the redshifts are at $z>1.2$; several of these objects were previously known to be likely high-z interlopers since their 24$\mu$m fluxes suggested the initial low photo-z to be wrong. For most galaxies the photo-z change is $\Delta z/(1+z) <0.02$ with some wiggles around the redshift diagonal and a scatter $<0.01$ around the wiggles. 

The restframe colours of galaxies are affected in two ways (see right panel of Fig.~\ref{zzcomp}): broadly the mean $U-V$ colour is now $0\fm11$ redder due to the overall recalibration. This first-order effect is independent of redshift since the calibration change of different filters is correlated by the monotonic calibration drift. Then the wiggles in photo-z change transform into wiggles in the colour change with redshift, and the scatter from old to new redshifts leads to scatter in the colour update.

An earlier comparison of the three COMBO-17 fields, CDFS, A901 and S11, showed that the colour-magnitude relation of red galaxies in the initial CDFS catalogue was $\sim 0\fm1$ bluer than that in the A901 and S11 field \citep{B04}. Now, the colour-magnitude relations in the three fields are consistent. The $U-V$ colour change leads to higher mass-to-light ratios in the galaxies: $\log M/L$ increases on average by 0.05~dex for blue galaxies and by 0.09~dex for red galaxies at $z=[0.2,1]$ according to the relation between $M/L$ and $B-V$ by Bell \& de Jong (2001). How that propagates into mass estimates depends on luminosity changes as well. At $z\approx 0.55$ the restframe V-band is centred on the observed-frame I-band, and there the new redshifts are lower by $\langle\Delta z\rangle\approx 0.015$, so reduced luminosity distance and increased I-band fluxes compensate each other. The net result for the restframe V-band luminosity is a dimming by $\sim 0\fm1$ at $z=0.2$, little change at $z=0.5$ and a brightening by $\sim 0\fm07$ at $z=0.8$. Hence, the resulting COMBO-17 galaxy masses on the ECDFS change very little for blue low-z galaxies, but rise by $\sim 10$\% and 20\% for blue and red galaxies at $z=0.5$, and by up to 30\% for red galaxies at $z=0.8$. We repeat, that the SEDs and masses in other COMBO-17 fields remain unaffected.

\begin{figure}
\centering
\includegraphics[clip,angle=270,width=0.9\hsize]{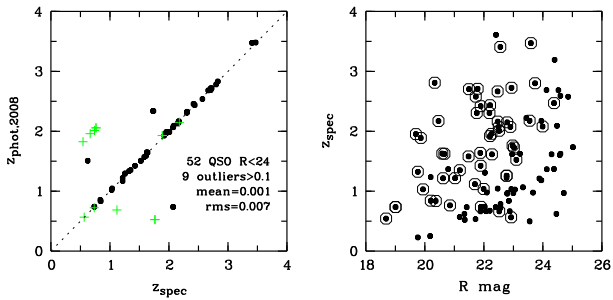}
\caption{{\it Left:} Photometric vs. spectroscopic QSO redshifts: excluding outliers the rms is 0.007 and the bias is consistent with zero. Crosses are objects with two ambiguous photo-z solutions and are responsible for most outliers {\it Right:} Type-1 AGN, encircled are those classed as QSO by COMBO-17: all but three QSOs with $M_B<-22$ and $z>0.5$ are found, while fainter AGN are dominated by the host galaxy SED. \label{speccompQ}}
\end{figure}

\subsection{QSOs}

The number and identity of objects classified as stars, white dwarfs, galaxies and QSOs are conserved at a $>95\%$ level. Fig.~\ref{speccompQ} shows type-1 AGN with spectroscopy from Szokoly et al. (2004) and Silverman et al. (in preparation). The right panel illustrates the QSO selection completeness in COMBO-17: encircled objects are classified as QSO by COMBO-17, so the selection is complete except in the regime of $M_B<-22$ and $z>0.5$ for three objects ($\sim 6\%$). 

The left panel shows photometric redshifts of the QSOs found by COMBO-17: Nine objects are outlying, in six of which the H$\beta$ and MgII emission lines are confused, perhaps because the continuum shapes in the data show variations not present in the template model. Six of the outliers have been flagged by the code as ambiguous and two alternative redshifts have been listed in the catalogue. In all six cases, the correct redshift was statistically less preferred by the code, while in only one of the ambiguous objects was the preferred redshift the correct one (cross on the diagonal). We remind users of the COMBO-17 catalogue that ambiguous redshifts are flagged by having alternative redshifts appear in the column $mc\_z2$.

Magnitude variability is accounted for in the SED fitting by using repeat R-band observations as a reference for the other filters, but potential colour variability is not accounted for. Finally: the non-outlying 43 QSOs have excellent accuracy, twice as good as in the initial release: $\delta z/(1+z)$ deviations of only 0.007 in rms and 0.001 bias.

\section{Conclusion}

The calibration of the CDFS field of COMBO-17 is updated over the original publication of the data in 2004. Each COMBO-17 field is calibrated by two spectrophotometric standard stars in each of its fields. While the calibration of the other COMBO-17 fields was straightforward, the two stars on the CDFS suggested calibrations that were inconsistent in colour at the 0.15~mag level from B to I. Both were marginally consistent with the colours of the Pickles atlas, so the choice was unconstrained. Wolf et al. (2004) ended up choosing the wrong star and introduced a colour bias to the blue. Here, we have changed the calibration to follow the other star, and it is now consistent with both the GaBoDS and MUSYC photometry. The consequences of the calibration change for the photometric redshifts is little when all 17 filters are used, but larger when only broad bands are used. Broad-band photo-z's hinge more on colours than on spectral features that are traced in medium-band photo-z's. However, a small positive change in the photo-z's is observed. Galaxies at $R<21$ and QSOs at $R<24$ have photo-z's accurate to within $<0.01$ rms of $\delta z/1(+z)$ after exclusion of outliers. The new catalogue version is accessible from the COMBO-17 or CDS websites.

\begin{acknowledgements}
CW was supported by an STFC Advanced Fellowship. HH by the European DUEL RTN, project MRTN-CT-2006-036133. We thank John Silverman for providing AGN redshifts prior to publication and Bret Lehmer for making his redshift compilation available electronically.
\end{acknowledgements}

\end{document}